ATOMS, SPECTRA,
AND RADIATIONS

# Experimental Implementation of a Four-Level N-type Scheme for the Observation of Electromagnetically Induced Transparency

V. M. Éntin\*, I. I. Ryabtsev, A. E. Boguslavskiĭ, and I. M. Beterov[†]

*Institute of Semiconductor Physics, Siberian Division, Russian Academy of Sciences,
pr. Akademika Lavrent'eva 13, Novosibirsk, 630090 Russia*
*\*e-mail: ventin@isp.nsc.ru*

**Abstract**—A nondegenerate four-level N-type scheme was experimentally implemented to observe electromagnetically induced transparency (EIT) at the $^{87}$Rb $D_2$ line. Radiations of two independent external-cavity semiconductor lasers were used in the experiment, the current of one of them being modulated at a frequency equal to the hyperfine-splitting frequency of the excited $5P_{3/2}$ level. In this case, apart from the main EIT dip corresponding to the two-photon Raman resonance in a three-level Λ-scheme, additional dips detuned from the main dip by a frequency equal to the frequency of the HF generator were observed in the absorption spectrum. These dips were due to an increase in the medium transparency at frequencies corresponding to the three-photon Raman resonances in four-level N-type schemes. The resonance shapes are analyzed as functions of generator frequency and magnetic field.© *2000 MAIK "Nauka/Interperiodica".*

The nonlinear interference effect [1] underlies numerous studies on coherent population trapping [2, 3], inversionless amplification [4], electromagnetically induced transparency [5], atomic laser cooling and capture [6], phase control of atomic states [7], etc. Initiation of electromagnetically induced transparency (EIT) in a medium with coherent pumping is one of the most prominent interference phenomena and manifests itself as narrow dips in the absorption spectra of atoms interacting with multifrequency resonance radiation.

As a rule, the experiments on EIT of alkali-metal atoms are carried out using the Λ-scheme of transitions in the laser radiation field modulated at a frequency equal to the ground-state hyperfine- or Zeeman-splitting frequency [8] or in the field of two phase-locked lasers [9]. The scanning of modulation frequency (or frequency difference in the case of two lasers) gives rise to a narrow interference resonance at the point corresponding to the exact two-photon Raman resonance between two ground-state sublevels coupled to the same excited state.

The use of the four-level N-type schemes of resonance transitions, for which Zeeman coherence can be spontaneously transferred from the excited to the ground level [10], opens up new opportunities for studying nonlinear phenomena. Such a scheme was investigated both theoretically [11] and experimentally [8, 12] for the quasidegenerate upper and lower states interacting with a two-frequency laser emission in a weak magnetic field. It was shown that the coherence transfer can give rise not only to the EIT but also to the electromagnetically induced absorption.

This paper reports the results of experimental implementation of the nondegenerate four-level N-type scheme for the observation of the EIT in a three-frequency laser radiation field. The experiments were carried out using the radiations of two independent semiconductor lasers, one of which was modulated at a frequency equal to the hyperfine-splitting frequency of the excited $5P_{3/2}$ state of the $^{87}$Rb atom. Therefore, the three-photon Raman resonance N-type scheme has been implemented for initiating EIT.

Experiments were carried out on the optical transitions between the hyperfine-structure components of the $^{87}$Rb $D_2$ absorption line (the $5S_{1/2} \longrightarrow 5P_{3/2}$ transition at a wavelength of 780 nm). The experimental setup is schematically shown in Fig. 1a. Two HL7851MG diode lasers with external cavities were used as radiation sources. The generation linewidth did not exceed 1 MHz. For tuning to different transitions, a portion of the radiation was diverted into two independent saturated-absorption sections *1* and *2* (according to [13]). The remaining portion of the radiation was combined in a collinear geometry. The wave fronts were carefully brought into coincidence by means of a set of mirrors and diaphragm *3*. The polarization plane of one of the radiations was turned through 90° by a half-wave plate for further detection of the separate absorption signals of the two beams using polaroids *4*. The beams passed through the absorption cell containing the $^{87}$Rb isotope vapor at temperatures from 20 to 40°C. The laser beam intensities were as high as 20 and 30 mW/cm$^2$ for lasers *1* and *2*, respectively.

[†] Deceased.

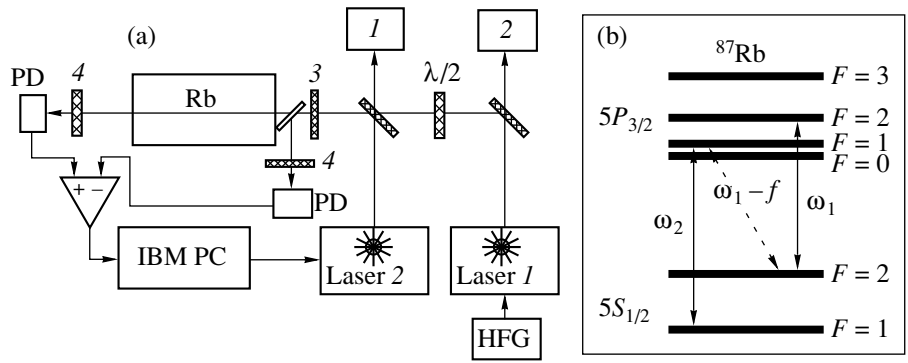

**Fig. 1.** (a) Experimental setup. (*1* and *2*, schemes of saturated absorption spectroscopy; *3*, diaphragm; *4*, polaroids; PD, photodiodes; HFG, high-frequency generator). (b) Scheme of the $^{87}$Rb $D_2$ line transitions occurring in the presence of HF modulation of laser *1*.

High-frequency modulation of the current of laser *1* was accomplished by an HF generator. Its frequency $f$ was varied near 156.9 MHz, i.e., near the frequency difference between the hyperfine-structure sublevels $F = 1$ and 2 of the excited $5P_{3/2}$ state (Fig. 1b). Switching on the generator resulted in the appearance of side frequencies shifted by $\pm f$ from the center frequency of the emission spectrum of laser *1*. Their intensities were one-tenth that of the center frequency.

The direction and strength of the constant magnetic field in the cell were defined by three pairs of Helmholtz coils. This made it possible to compensate the laboratory magnetic field, which was equal to 0.7 G and considerably affected the resonance amplitudes.

Absorption of the radiation from laser *1* was recorded simultaneously by tuning the frequency $\omega_2$ of laser *2* within the Doppler profile of a group of the $5S_{1/2}(F = 1) \longrightarrow 5P_{3/2}(F = 0, 1, 2)$ transitions (Fig. 1b), Doppler broadening of individual resonances was 520 MHz. In this case, the center frequency $\omega_1$ of laser *1* was fixed and tuned to the center of the Doppler profile of a group of the $5S_{1/2}(F = 2) \longrightarrow 5P_{3/2}(F = 1, 2, 3)$ transitions. The frequency of laser *2* was computer-controlled.

The absorption signal $\omega_1$ represented the sum of signals from three groups of atoms with different velocity projections. Due to the Doppler effect, each of them was in resonance with one of the three transitions. In the absence of emission $\omega_2$, the absorption at frequency $\omega_1$ was at a certain fixed level (about 10%) determined by the atomic concentration, saturation parameters, relaxation rates, etc.

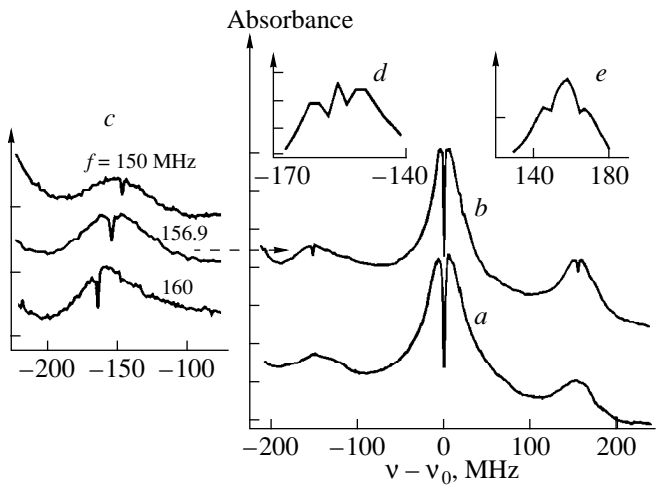

**Fig. 2.** (*a*) Absorption of the radiation of laser *1* recorded upon scanning the frequency of laser *2*. (*b*) The same spectrum with the switched-on HF modulation of laser *1*. (*c*) Positions of additional resonances at different frequencies of HF generator. (*d*) Shape of the additional resonance $5S_{1/2}(F = 1) \longleftrightarrow 5P_{3/2}(F = 1) \longleftrightarrow 5S_{1/2}(F = 2) \longleftrightarrow 5P_{3/2}(F = 2)$ (left in Fig. 2b) in a transverse magnetic field of 10 G. (*e*) The same for the $5S_{1/2}(F = 1) \longleftrightarrow 5P_{3/2}(F = 2) \longleftrightarrow 5S_{1/2}(F = 2) \longleftrightarrow 5P_{3/2}(F = 1)$ resonance (right in Fig. 2b).

After switching on and scanning frequency $\omega_2$, several peaks with different amplitudes appeared in the absorption spectrum $\omega_1$ (Fig. 2a). The increase in the absorption at certain values of frequency $\omega_2$ was caused by an increase in the population of the $5S_{1/2}(F = 2)$ level due to the spontaneous decay of the $5P_{3/2}(F = 1, 2)$ levels excited by radiation $\omega_2$. The peak widths (ca. 40 MHz) were mostly due to the power broadening of the resonances and decreased with decreasing laser radiation intensity. The frequency spacings between the peaks were equal to the hyperfine-splitting frequencies of the excited $5P_{3/2}$ state.

In the absence of HF modulation of laser *1*, only one of the peaks had the interference EIT dip at its center (Fig. 2a). The dip corresponded to the summarized signal from the two-photon Raman resonances $5S_{1/2}(F = 1) \longleftrightarrow 5P_{3/2}(F = 1) \longleftrightarrow 5S_{1/2}(F = 2)$ and $5S_{1/2}$ ($F = 1$) ⟷ $5P_{3/2}$ ($F = 2$) ⟷ $5S_{1/2}$ ($F = 2$) in two groups of atoms with different velocity projections. The dip width fell in the range 2.6–4 MHz and was less than the natural linewidth (6 MHz). It is known that the width of EIT resonances depends strongly on the fluctuations of the phase and frequency differences of laser radiations and can be tangibly reduced upon matching the phases of two radiations if the depolarizing collisions are absent [10]. In our case, two independent lasers and a cell with rubidium vapor at low pressure without buffer gas were used, so that the width of the interference resonances was mainly determined by the width of the laser spectra.

The EIT resonance was most contrasting in the case of orthogonal linear polarizations of radiations $\omega_1$ and $\omega_2$. To enhance contrast, it was necessary to compensate the laboratory magnetic field.

Switching on the HF modulation of laser 1 resulted in the appearance of the additional EIT dips at the tops of the two neighboring absorption peaks detuned by the $F = 1, 2$ interval of the hyperfine structure of the $5P_{3/2}$ state (Fig. 2b). Because of the presence of side frequencies in the spectrum of laser 1, these dips were due to the three-photon Raman resonances $5S_{1/2}$ ($F = 1$) ⟷ $5P_{3/2}$ ($F = 1$) ⟷ $5S_{1/2}$ ($F = 2$) ⟷ $5P_{3/2}$ ($F = 2$) (left dip in Fig. 2b) and $5S_{1/2}$ ($F = 1$) ⟷ $5P_{3/2}$ ($F = 2$) ⟷ $5S_{1/2}$ ($F = 2$) ⟷ $5P_{3/2}$ ($F = 1$) (right dip in Fig. 2b) in two groups of atoms with different velocity projections. The frequency detuning of the additional dips from the main EIP resonance was equal to the HF-generator frequency. As the frequency varied over the range 156.9 ± 15 MHz, the dips changed their positions at the tops of the peaks (Fig. 2c) and disappeared at large detuning.

Note that, at the exact resonance, the radiations at three frequencies, $\omega_1$, $\omega_2$, and ($\omega_1 + f$) or ($\omega_1 - f$), interact simultaneously with each atomic group. This interaction generates a coherent superposition of atomic state, for which the absorption of laser radiations decreases (dark resonance). The appearance of the dark resonance underlies the phenomenon of electromagnetically induced transparency.

The additional EIT dips were most contrasting in the presence of a weak (≤0.5 G) longitudinal (collinear with the direction of the laser beams) magnetic field but disappeared at a field strength above 2.5 G. At the same time, the indicated fields had almost no effect on the shape and amplitude of the main EIT resonance.

In the case of a transverse magnetic field, the resonances did not disappear, allowing their Zeeman splitting to be observed. Figures 2d and 2e show the experimental records of the additional dips in the presence of a 10-G magnetic field aligned with the polarization vector of the $\omega_1$ wave. A sizably different behavior of the resonances in Figs. 2d and 2e is noteworthy. Despite the low contrast, the resonance structure shown in Fig. 2e allows one to assume that an inverted dip is present at the center and that its origin is the same as the origin of the inverted dips observed in [8, 11, 12]; i.e., it is the electromagnetically induced absorption. However, an analysis of the rather complex shape of the resonances observed in our case requires careful theoretical consideration and numerical calculations with taking into account the Zeeman splitting of atomic levels.

The effect demonstrated in this work is the implementation of the nondegenerate four-level N-type scheme for initiating the EIT in the field of three-frequency laser radiation. At the same time, it can be treated as a frequency shift of the EIT resonances under the HF modulation of the laser radiation. This effect can be applied in a wide variety of problems in the field of nonlinear laser spectroscopy, in particular, in various schemes of laser cooling and in the experiments on inversionless amplification and lasing. Note in conclusion that the observed resonances can have extremely small widths if the magnetic field is accurately compensated and the laser radiations are closely phase-locked.

## ACKNOWLEDGMENTS

We are grateful to A. M. Tumaĭkin, V. I. Yudin, and A. V. Taĭchenachev for assistance and discussion of the experimental results. This work was supported by the Russian Foundation for Basic Research, project nos. 97-02-18551, 99-02-17131, and 00-02-17924.

*Translated by R. Tyapaev*